\documentclass[twocolumn,conference]{IEEEtran}
\usepackage{float}
\usepackage{amsmath}
\usepackage{amssymb}
\usepackage{graphicx}
\usepackage[unicode=true,
 bookmarks=true,bookmarksnumbered=true,bookmarksopen=true,bookmarksopenlevel=1,
 breaklinks=false,pdfborder={0 0 0},pdfborderstyle={},backref=false,colorlinks=false]
 {hyperref}
\hypersetup{pdftitle={Your Title},
 pdfauthor={Your Name},
 pdfpagelayout=OneColumn, pdfnewwindow=true, pdfstartview=XYZ, plainpages=false}

\makeatletter

\DeclareFontEncoding{LGR}{}{}
\DeclareRobustCommand{\greektext}{%
  \fontencoding{LGR}\selectfont\def\encodingdefault{LGR}}
\DeclareRobustCommand{\textgreek}[1]{\leavevmode{\greektext #1}}
\ProvideTextCommand{\~}{LGR}[1]{\char126#1}

\providecommand{\tabularnewline}{\\}

\usepackage[caption=false,font=footnotesize]{subfig}

\makeatother

\begin{document}
\IEEEoverridecommandlockouts

\hypersetup{bookmarks={false}}

\title{Improved linear direct solution for asynchronous radio network localization
(RNL)}

\author{\IEEEauthorblockN{Juri~Sidorenko, Norbert~Scherer-Negenborn, Michael~Arens, Eckart
Michaelsen }\IEEEauthorblockA{Fraunhofer Institute of Optronics, System Technologies and Image Exploitation
IOSB\\
Gutleuthausstrasse 1, 76275 Ettlingen, Germany. \\
Juri.Sidorenko@iosb.fraunhofer.de}}
\maketitle
\begin{abstract}
In the field of localization the linear least square solution is frequently
used. This solution is compared to nonlinear solvers more effected
by noise, but able to provide a position estimation without the knowledge
of any starting condition. The linear least square solution is able
to minimize Gaussian noise by solving an overdetermined equation with
the Moore\textendash Penrose pseudoinverse. Unfortunately this solution
fails if it comes to non Gaussian noise. This publication presents
a direct solution which is able to use pre-filtered data for the LPM
(RNL) equation. The used input for the linear position estimation
will not be the raw data but over the time filtered data, for this
reason this solution will be called direct solution. It will be shown
that the presented symmetrical direct solution is superior to non
symmetrical direct solution and especially to the not pre-filtered
linear least square solution.

\end{abstract}

\IEEEpeerreviewmaketitle{}

\section{Introduction}

Radio network-based localization is a radio wave-based positioning
method, whereby a set of sensors at known positions (base stations)
estimate the unknown sensor position (M). Range measurement can be
accomplished using different principles, such as 'Round Trip Time
of Flight (RTT, RTToF)'. With this approach, the base stations send
the signal and it is sent back by M. This technique is very similar
to radar, hence every measurement can be referred to as 'Time Of Arrival
(TOA) '. Alternatively, the base stations can be passive and only
the M emits the signal, referred to as 'Local Position Measurement
(LPM)'. The base stations have to be synchronized, which can be achieved
with a second transponder (T) at a known position. In contrast to
the elliptical TDOA method \cite{Zhou2011}, M and T do not communicate
with each other, thus every measurement has a time offset. The unknown
sensor position can be estimated using the direct linear (closed loop)
solution instead of Taylor-series expansion \cite{FOY1976} or nonlinear
solvers. The direct solution has the advantage that no starting conditions
are required compared to the nonlinear or Taylor solution. However,
every real measurement is affected by noise, which is in the best
Gaussian case. Unfortunately, reflections and other non-Gaussian disturbances
can also be found in the measurements. The best approach is to filter
this data before lateration, since even raw data with Gaussian noise
can become non-Gaussian after a non-linear operation, caused by nonlinear
optimization. This measurement principal requires data transformation
to eliminate the time offset, filter the data and use it as an input
for the linear solution. The Inmotiotec LPM system \cite{Pourvoyeur2005,Resch2012}
is an example of this kind of system and is able to provide an update
rate of 1000 Hz with high three-dimensional position accuracy. 

\section{Previous Work}

A linear algebraic solution (direct solution) is frequently used in
the field of position estimation. One of those most commonly used
is Bancroft\textquoteright s method \cite{Bancroft1985}, which is
well analyzed and described in \cite{Abel1991,Chaffee1994}. Furthermore,
it has been shown that the nonlinear solvers, such as the Gauss\textendash Newton
algorithm, provide more accurate solutions for overdetermined cases
than the linear solution solved by the Moore\textendash Penrose pseudoinverse
\cite{Sirola2010}. In paper \cite{Sidorenko2016} we show an approach
to using filtered measurements for the linear direct solution of the
TOA-LPM equation. This solution is not symmetrical and less numerically
stable, therefore an improved solution will be presented in this work.
The Abatec LPM system itself is well described in \cite{Pourvoyeur2005,Resch2012}.
Previous publications about the Abatec LPM are mainly based on measurement
principles \cite{Pourvoyeur2005,Resch2012} and how the sensor data
can be fused and filtered to detect outliers \cite{R.Pfeil2009} and
obtain the most accurate position \cite{Pourvoyeur2006a} after multilateration.
The latest publications on LPM focus on the numerical solvers. In
general, LPM uses a Bancroft algorithm \cite{Pourvoyeur2006,Stelzer2004,R.Pfeil2009}
to estimate the position of the transponder. 

\section{Methodology }

The general LPM equation is

\begin{equation}
R_{i}=O+||M-B_{i}||-||T-B_{i}||\label{eq:LPM}
\end{equation}

The pseudo range measurement (R), consists of the flight time between
the first transponder and one base station subtracted from the reference
transponder to the same base station. In this case, every time offset
(O=time offset{*}speed of light) remains the same for every base station
at one measurement but changes rapidly over time. The Euclidian distance
between the reference station equates with T indices eq.(\ref{eq:2})
and transponder with M indices eq.(\ref{eq:3}) to the base stations
B with indices number ($i=1\rightarrow n$ ) .

\begin{equation}
||T-B_{i}||=\sqrt{(x_{i}-x_{T})^{2}+(y_{i}-y_{T})^{2}}\label{eq:2}
\end{equation}

\begin{equation}
||M-B_{i}||=\sqrt{(x_{i}-x_{M})^{2}+(y_{i}-y_{M})^{2}}\label{eq:3}
\end{equation}

The coordinates of the base stations and reference station are known.
Only the transponder position and the offset have to be estimated.

\subsection{Not symmetrical direct solution}

In contrast to approaches such as Taylor-series expansion \cite{FOY1976},
for which starting information for the unknown variables is required,
it is possible to obtain the linear components x,y,z of the transponder
without derivation. The main LPM equation (\ref{eq:LPM}) can be simplified
by adding the known reference transponder range to the measurement
term $R$ (pseudo range).

\begin{equation}
L_{i}=R_{i}+||T-B_{i}||
\end{equation}

\begin{equation}
||M-B_{j}||^{2}-||M-B{}_{i}||^{2}=(L_{j}-O)^{2}-(L_{i}-O)^{2}
\end{equation}

The known quadratic terms of the transponder are eliminated, hence
the linear solution for the transponder position at the known base
station and reference station position is

\[
(\overrightarrow{B_{i}}-\overrightarrow{B_{j}})\cdotp\overrightarrow{M}-(L_{i}-L_{j})\cdot O=
\]

\begin{equation}
=\frac{1}{2}((\overrightarrow{B_{i}}{}^{2}-\overrightarrow{B_{j}}{}^{2})-(L_{i}^{2}-L_{j}^{2}))
\end{equation}

With:
\begin{center}
\begin{tabular}{ccc}
$\overrightarrow{M}=\left(\begin{array}{c}
x_{M}\\
y_{M}
\end{array}\right)$ &  & $\overrightarrow{B}=\left(\begin{array}{c}
x_{B}\\
y_{B}
\end{array}\right)$\tabularnewline
\end{tabular} 
\par\end{center}

In \cite{Sidorenko2016} it is shown that the linear solution for
the LPM is highly affected by noise. The measurement $L_{1}$ cannot
be filtered as the time offset change with respect to time is 1000
times higher than the range change itself. Therefore, the offset is
eliminated by subtracting one base station measurement from the other
$(L_{i}-L_{j})$. In the next step this data is filtered over time.
At this point it does not matter what kind of filter is used, it is
only important that filtering takes place before position estimation
and that the filter uses the measurement difference $(L_{i}-L_{j})$
as an input. For the following calculations we only assume that for
every measurement we already have difference $(L_{i}-L_{j})$ the
filtered values F$(L_{i}-L_{j})$, the main aim is to use the filtered
values instead of the measurement differences between the base stations.
The $(L_{i}^{2}-L_{j}^{2})$ term is nonlinear but the filtered values
consist of the linear difference between the measurement ranges. One
solution to using the filtered values would be to make every base
station dependent on the same measurement error term \textgreek{a}i.
Every measurement is corrupted by the measurement error \textgreek{a}i,
hence the real measurement can be written as $L=\tilde{L}_{i}+\alpha$.
The connection between the measurement errors $\alpha_{i}$ and $\alpha_{j}$
can be found if the unfiltered measurement difference $((L_{i}+\alpha_{i})-(L_{j}+\alpha_{j}))$
is subtracted from the filtered values $F(L_{i}-L_{j})$. \\

\begin{equation}
F_{ij}=((\tilde{L}_{i}+\alpha_{i})-(\tilde{L}_{j}+\alpha_{j}))-F(L_{i}-L_{j})
\end{equation}

\begin{equation}
(\tilde{L}_{i}-\tilde{L}_{j})\thickapprox F(L_{i}-L_{j})
\end{equation}

The assumption that the noise can be neglected after the filtering,
this leads to the term $F_{ij}$ being the difference between the
noises of both signals.

\begin{equation}
F_{ij}=\alpha_{i}-\alpha_{j}
\end{equation}

\begin{equation}
\alpha_{j}=-F_{ij}+\alpha_{i}
\end{equation}

The measurement error $\alpha_{j}$is replaced by $-F_{ij}+\alpha_{i}$

\[
(\overrightarrow{B_{i}}-\overrightarrow{B_{j}})\cdotp\overrightarrow{M}-(\tilde{L}_{i}-\tilde{L_{j}}+F_{ij})\cdotp O=
\]

\[
=\frac{1}{2}((\overrightarrow{B_{i}}{}^{2}-\overrightarrow{B_{j}}{}^{2})-\tilde{L_{i}}^{2}-\tilde{L_{j}}^{2}-F_{ij}^{2}+
\]

\begin{equation}
+2\cdotp\alpha_{k}(\tilde{L}_{i}-\tilde{L_{j}}+F_{ij}))
\end{equation}

It can be observed that the time offset Z, depends on the same parameter
as the measurement error

\[
(\overrightarrow{B_{i}}-\overrightarrow{B_{j}})\cdotp\overrightarrow{M}-(\tilde{L}_{i}-\tilde{L_{j}}+F_{ij})\cdotp(O+\alpha_{k})=
\]

\begin{equation}
=\frac{1}{2}((\overrightarrow{B_{i}}{}^{2}-\overrightarrow{B_{j}}{}^{2})-\tilde{L_{i}}^{2}-\tilde{L_{j}}^{2}-F_{ij}^{2})
\end{equation}

With at least four base stations, the unknown coordinates of the transponder
can be estimated. With the filtered values the linear direct solution
provides better results, than with the unfiltered equation.

\begin{equation}
Ax=b
\end{equation}

This equation can be solved as: 

\begin{equation}
\left(\begin{array}{c}
x_{M}\\
y_{M}\\
O
\end{array}\right)=(A^{T}*A)^{-1}A^{T}*b\label{eq:Lsolver}
\end{equation}

\subsection{Symmetrical direct solution}

The main disadvantage of the previous equation is that every base
station measurement depends on one and the same base station, hence
the solution is not symmetrical. Therefore, a more robust approach
whereby every base station is used will be presented in the next part.
In contrast to the solution presented in 3.1, the nonlinear difference
between the measurements will be rewritten as. 

\begin{equation}
(L_{i}^{2}-L_{j}^{2})=(L_{i}-L_{j})\cdotp(L_{i}+L_{j})
\end{equation}

This leads to the term

\[
(\overrightarrow{B_{i}}-\overrightarrow{B_{j}})\cdotp\overrightarrow{M}-(L_{i}-L_{j})\cdotp(O-\frac{L_{i}+L_{j}}{2})=
\]

\begin{equation}
=\frac{1}{2}(\overrightarrow{B_{i}}{}^{2}-\overrightarrow{B_{j}}{}^{2})
\end{equation}

The difference between two measurements $\varDelta_{ij}$can be replaced
by results of the filter.

\begin{equation}
\varDelta_{ij}=(L_{i}-L_{j})
\end{equation}

Filtering uses the differences between two measurements, as offset
O is equal for the same measurement for every base station. Therefore,
every measurement difference $(L-L_{j})$ can be replaced by the filtered
values. Only the sum between two measurements $(L_{1}+L_{j})$ is
unknown.

\begin{equation}
(\overrightarrow{B_{i}}-\overrightarrow{B_{j}})\cdotp\overrightarrow{M}-\varDelta_{i,j}\cdotp(O-\frac{L_{i}+L_{j}}{2})=\frac{1}{2}(\overrightarrow{B_{i}}{}^{2}-\overrightarrow{B_{j}}{}^{2})
\end{equation}

\begin{equation}
\varDelta_{ji}=-\varDelta_{ij},\ \varDelta_{ii}=0
\end{equation}

In the following example, it will be shown how the sum of two measurements
$(L_{i}+L_{j})$ is represented by the Differences of two measurements
$(L_{i}-L_{j})$. 

\subsection*{Symmetrical direct solution: Example with 5 base stations}

For five base stations the filtered measurement differences required
are

\begin{center}
\begin{tabular}{|c|c|c|c|c|}
\hline 
 & $BS_{1-j}$ & $BS_{2-j}$ & $BS_{3-j}$ & $BS_{4-j}$\tabularnewline
\hline 
j=2 & 1-2 & 2-3 & 3-4 & 4-5\tabularnewline
\hline 
j=3 & 1-3 & 2-4 & 3-5 & \tabularnewline
\hline 
j=4 & 1-4 & 2-5 &  & \tabularnewline
\hline 
j=5 & 1-5 &  &  & \tabularnewline
\hline 
\end{tabular}
\par\end{center}

The sum of measurements $L_{i}+L_{j}$ for five base stations can
be represented by the filtered measurement differences:

\begin{center}
\begin{tabular}{c}
$L_{1}+L_{3}=(L_{1}+L_{2})-(\varDelta_{23})$\tabularnewline
$L_{1}+L_{4}=(L_{1}+L_{2})-(\varDelta_{24})$\tabularnewline
$L_{1}+L_{5}=(L_{1}+L_{2})-(\varDelta_{25})$\tabularnewline
$L_{2}+L_{3}=(L_{1}+L_{2})-(\varDelta_{13})$\tabularnewline
$L_{2}+L_{4}=(L_{1}+L_{2})-(\varDelta_{14})$\tabularnewline
$L_{2}+L_{5}=(L_{1}+L_{2})-(\varDelta_{15})$\tabularnewline
$L_{3}+L_{4}=(L_{1}+L_{2})-(\varDelta_{13})-(\varDelta_{24})$\tabularnewline
$L_{3}+L_{5}=(L_{1}+L_{2})-(\varDelta_{13})-(\varDelta_{25})$\tabularnewline
$L_{4}+L_{5}=(L_{1}+L_{2})-(\varDelta_{25})-(\varDelta_{34})$\tabularnewline
\end{tabular}
\par\end{center}

Now every base station depents on the unkown sum component $(L_{1}+L_{2})$.
Insted of $(L_{1}+L_{2})$ it is also possible to use any combination
of ($L_{i}+L_{j}$ ). Our aim is to make the equation equally dependent
on all the base stations and not only on fixed base station combinations
($L_{i}+L_{j}$ ). For this reason, the sum ($L_{i}+L_{j}$) between
two base stations will be replaced by the sum of all base stations.

If we stay by the example with five base stations, the sum S would
be.

\begin{equation}
S=\sum_{i=1}^{n}L_{i}
\end{equation}

This unkown sum S should now fit for every ($L_{i}+L_{j}$). If the
indizes are i=1 and j=2, the sum S need to be transfomred in such
a way that $L_{3}+L_{4}+L_{5}$ are eliminated by only using the differences
between the $L_{i}$and $L_{j}$. For
\begin{equation}
L_{1}+L_{2}\neq S+(L_{1}-L_{3})+(L_{1}-L_{4})+(L_{1}-L_{5})
\end{equation}
the components $L_{3}$,$L_{4}$ and$L_{5}$are eliminated but now
$L_{1}$is represented four times instead of once. If the equation
is changed to
\[
L_{1}+L_{2}\neq S+(L_{1}-L_{3})+(L_{1}-L_{4})+(L_{1}-L_{5})+
\]

\begin{equation}
+(L_{2}-L_{3})+(L_{2}-L_{4})+(L_{2}-L_{5})
\end{equation}

the measurements $L_{1}$ and $L_{2}$ are now overrepresented four
times and $L_{3}$, $L_{4}$ and $L_{5}$ are overrepresented twice.
By multiplying the measurement differences by 0.5 and adding them
to the sum S, the $L_{3}$, $L_{4}$ and $L_{5}$ are eliminated but
$L_{1}$ and $L_{2}$ now have the factor 5/2 instead of one. The
numerator represents the number of base stations used, in this example
five. The term $(L_{i}+L-j)$ can now be expressed as

\[
L_{1}+L_{2}=\frac{2}{5}\cdotp(S+\frac{1}{2}((L_{1}-L_{3})+(L_{1}-L_{4})+(L_{1}-L_{5})+
\]

\begin{equation}
+(L_{2}-L_{3})+(L_{2}-L_{4})+(L_{2}-L_{5})))
\end{equation}
The general equation for every $L_{i}+L_{j}$ becomes

\begin{equation}
L_{i}+L_{j}=\frac{2}{n}\cdotp S+\frac{1}{n\cdotp2}\cdotp(\sum_{k\neq i,j}(\varDelta_{ik})+\sum_{k\neq i,j}(\varDelta_{jk}))
\end{equation}

with the variable n, which stands for the number of base stations.
The sum of the measurements between two base stations can now be replaced
by the following term, whereby every other base station is used equally.

\begin{equation}
\frac{L_{i}+L_{j}}{2}=\frac{1}{n}\cdotp S+\frac{1}{n\cdotp2}\cdotp(\sum_{k\neq i,j}\varDelta_{ik}+\sum_{k\neq i,j}\varDelta_{jk})
\end{equation}

The final symmetrical direct solution, with the unkown variables $x_{MT}$,$y_{MT}$,$z_{MT}$,$O$
and$S$ equates:

\begin{center}
\[
(\overrightarrow{B_{i}}-\overrightarrow{B_{j}})\cdotp\overrightarrow{M}-\varDelta_{i,j}\cdotp(O-\frac{1}{n}\cdotp S)=
\]
\par\end{center}

\begin{center}
\begin{equation}
=\frac{1}{2}(\overrightarrow{B_{i}}{}^{2}-\overrightarrow{B_{j}}{}^{2})-\varDelta_{i,j}\cdotp\frac{1}{2\cdotp n}\cdotp(\sum_{k\neq i,j}\varDelta_{ik}+\sum_{k\neq i,j}\varDelta_{jk})
\end{equation}
\par\end{center}

\section{Results}

In the following two methods (symmetrical and non-symmetrical) filtering
the direct solution will be compared. The five base station positions
are located on a circle with a radius of 10 metres and the transponder
position measurement is calculated for every metre in the $60\ m^{2}$
square area. Furthermore, the measurements have been corrupted by
Gaussian noise with a variance of $0.064\ m^{2}$. This noise represents
the filtering error not the measurement noise. 

\begin{table}[H]
\begin{centering}
\begin{tabular}{|c|c|c|c|c|c|c|}
\hline 
Base station & 1 & 2 & 3 & 4 & 5 & 6\tabularnewline
\hline 
\hline 
X-Axis {[}m{]} & 10 & 5 & -5 & -10 & -5 & 5\tabularnewline
\hline 
Y-Axis {[}m{]} & 0 & 8.66 & 8.66 & 0 & -8.66 & -8.66\tabularnewline
\hline 
\end{tabular}\ \\
\ 
\par\end{centering}
\caption{Base station position \label{tab:Base-station-position}}
\end{table}

\begin{figure}[H]
\begin{centering}
\includegraphics[scale=0.5]{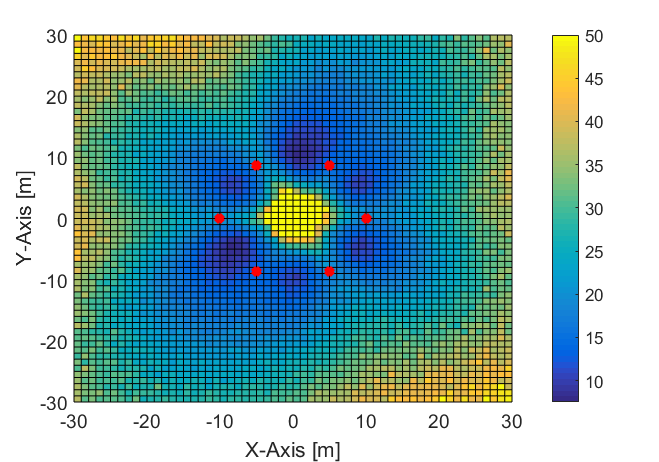}
\par\end{centering}
\caption{Matrix condition with BS 1 as reference. \protect \\
Colours from yellow to blue: condition at the specific position. The
red dots: base station positions.\label{fig:Matrix-condition-previous}. }
\end{figure}

\begin{figure}[H]
\begin{centering}
\includegraphics[scale=0.5]{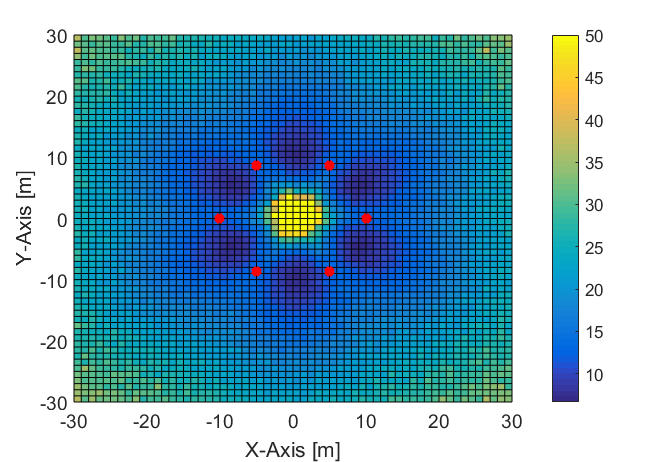}
\par\end{centering}
\caption{Matrix condition symmetric approach.\protect \\
Colours from yellow to blue: condition at the specific position. The
red dots: base station positions..\label{fig:Matrix-condition-new}}
\end{figure}

The position error of the previous method is subtracted from the new
one at any position in the square area. Positive error differences
indicate that the error with the second method is smaller. On the
other hand, negative error difference shows that the error with the
second approach is higher compared to the non-symmetrical solution.
In the set-up presented, 56.11\% have a positive error difference
and 43.88\% a negative one. Therefore, the second approach is 13\%
superior to the first one. In some test scenarios, where the geometrical
constellation of the base stations is difficult for the lateration
of the transponder position, the difference between the non-symmetrical
and symmetrical approach increases by 30\%. The first approach (non-symmetrical)
always uses the same base station (base station one) from which the
others are subtracted. If this transformation station is selected
by the best condition of the coefficient matrix $||A||*||A^{-1}||$,
the error difference between the new and previous approach is almost
always a value between 50.44 \% and 49.55 \%. The increase in noise
when selecting difficult geometrical constellations of the base stations
leads to a higher difference between the symmetrical and non-symmetrical
approach, with better results for the symmetrical approach. The condition
of the coefficient matrix at different transponder positions can be
seen in figure \ref{fig:Matrix-condition-previous} and \ref{fig:Matrix-condition-new}.

It can be observed that the coefficient matrix condition with the
previous method, figure \ref{fig:Matrix-condition-previous} is not
symmetrical compared to the new approach figure \ref{fig:Matrix-condition-new}.
Base station selection with the best condition slightly changes the
results but still underlies the symmetrical approach. The condition
appears to be best in the area between the base stations. This is
not always the case and depends on the base station constellation.

\section{Conclusion}

'LPM' is a nonlinear offset corrupted equation, whereby any transformation
to a linear solution leads to a high noise impact on the outcome.
We present a new numerically more stable direct solution, which is
able to work with prefiltered data. In contrast to a non-filtered
linear least square solution, this filtered direct solution is statically
not correct, as not the data corrupted by Gaussian noise is used but
the output of the filter. The results of this filtered direct solution
are less influenced by the noise and therefore more suitable for use
as starting values for the nonlinear solver. Furthermore, the symmetrical
approach does not require a specific base station that is used for
filtering, but instead all base stations play an equal part in finding
a solution. The results of the new approach are at least 10 \% better,
compared to approaches whereby only the same base station is used.
The best result for the first solution appears if the reference station
that causes the best condition matrix is selected. In the new solution,
there is no need to find the base station which causes the best matrix
condition since every base station is used equally. Especially when
the noise increases or the geometrical set-up is unfavourable, the
results of the symmetrical approach are superior to its non-symmetrical
counterpart.

\bibliographystyle{plain}
\bibliography{bib_LMP}

\end{document}